\documentclass[12pt]{elsarticle}
\usepackage{epsfig}

\begin{document}
\begin{frontmatter}

\title{Radioactive contamination of SrI$_2$(Eu) crystal scintillator}

\author[INFN-Roma2] {P.~Belli}
\author[INFN-Roma2,Univ-Roma2]{R.~Bernabei\corref{cor1}}
 \cortext[cor1]{Corresponding author. E-mail:
 rita.bernabei@roma2.infn.it}
\author[LNGS] {R.~Cerulli}
\author[KINR] {F.A.~Danevich}
\author[ISMA] {E.~Galenin}
\author[ISMA] {A.~Gektin}
\author[INFN-Roma1,Univ-Roma1] {A.~Incicchitti}
\author[NAUKMA] {V.~Isaienko}
\author[KINR] {V.V.~Kobychev}
\author[LNGS] {M.~Laubenstein}
\author[KINR] {S.S.~Nagorny}
\author[KINR] {R.B.~Podviyanuk}
\author[ISMA] {S.~Tkachenko}
\author[KINR] {V.I.~Tretyak}

\address[INFN-Roma2]{INFN sezione Roma ``Tor Vergata'', I-00133 Rome, Italy}
\address[Univ-Roma2]{Dipartimento di Fisica, Universit\`a di Roma ``Tor Vergata'', I-00133, Rome, Italy}
\address[LNGS]{INFN, Laboratori Nazionali del Gran Sasso, I-67100 Assergi (AQ), Italy}
\address[KINR]{Institute for Nuclear Research, MSP 03680 Kyiv, Ukraine}
\address[ISMA] {Institute for Scintillation Materials, 61001, Kharkiv, Ukraine}
\address[INFN-Roma1]{INFN sezione Roma, I-00185 Rome, Italy}
\address[Univ-Roma1]{Dipartimento di Fisica, Universit\`a di Roma ``La Sapienza'', I-00185 Rome, Italy}
\address[NAUKMA] {National University of Kyiv-Mohyla Academy, 04655 Kyiv, Ukraine}

\begin{abstract}

A strontium iodide crystal doped by europium (SrI$_2$(Eu)) was
produced by using the Stockbarger growth technique. The crystal was
subjected to a characterization that includes relative photoelectron
output and energy resolution for $\gamma$ quanta. The intrinsic
radioactivity of the SrI$_2$(Eu) crystal scintillator was tested both
by using it as scintillator at sea level and by ultra-low background HPGe
$\gamma$ spectrometry deep underground. The response of the SrI$_2$(Eu) detector to
$\alpha$ particles ($\alpha/\beta$ ratio and pulse shape) was
estimated by analysing the $^{226}$Ra internal trace contamination of the
crystal.
We have measured: $\alpha/\beta=0.55$ at $E_\alpha=7.7$ MeV, and no
difference in the time decay of the scintillation pulses induced by $\alpha$ particles and
$\gamma$ quanta. The application of the obtained results in the
search for the double electron capture and electron capture with
positron emission in $^{84}$Sr has been investigated at a level
of sensitivity: $T_{1/2}\sim 10^{15}-10^{16}$ yr. The results
of these studies demonstrate the potentiality of this material for a
variety of scintillation applications, including low-level counting
experiments.

\end{abstract}

\begin{keyword}

SrI$_2$(Eu) crystal scintillator \sep Radioactive contamination
\sep Double beta decay

\vskip 0.2cm

\PACS 29.40.Mc Scintillation detectors \sep 23.40.-s $\beta$
decay; double $\beta$ decay; electron and muon capture

\end{keyword}

\end{frontmatter}

\section{Introduction}

The strontium iodide was discovered as scintillator by Hofstadter in
1968 \cite{Hofstadter:1968}. The interest in this material increased
in the last few years because of the high light output ($> 100~000$ photons/MeV)
and of the good energy resolution ($\approx3\%$ at 662 keV),
recently reported in refs. \cite{Cherepy:2008,Cherepy:2009a,Loef:2009}. The main
properties of SrI$_2$(Eu) crystal scintillators are presented in
Table \ref{properties}.

\nopagebreak
\begin{table}[!h]
 \caption{Properties of SrI$_2$(Eu) crystal scintillators}
\begin{center}
\begin{tabular}{lll}

 \hline
 Property                               & Value                 & Reference \\
  ~                                     & ~                     & ~         \\
  \hline

 Density (g/cm$^3$)                     & $4.5-4.6$             & \cite{Cherepy:2008,Loef:2009,Alekhin:2011} \\

 \hline
 Melting point ($^\circ$C)              & 515                   & \cite{Cherepy:2008}     \\

 \hline
  Structural type                       & Orthorhombic          & \cite{Cherepy:2008} \\

 \hline

 Index of refraction                    & $1.85$                & \cite{Tan:2011} \\

 \hline
 Wavelength of emission                 & ~                     & ~ \\
 maximum (nm)                           & $429-436$             & \cite{Cherepy:2008,Loef:2009,Alekhin:2011} \\

 \hline
 Light yield (photons/MeV)              & $(68-120)\times10^3$  & \cite{Cherepy:2008,Loef:2009,Glodo:2010}\\

 \hline
 Energy resolution (FWHM, \%)           & ~                     & ~ \\
 for 662 keV $\gamma$ of  $^{137}$Cs    & $2.6-3.7$             & \cite{Cherepy:2008,Loef:2009,Glodo:2010,Sturm:2011,Cherepy:2009b} \\

 \hline
 Scintillation decay time ($\mu$s)      & ~                     & ~ \\
 under X ray / $\gamma$ ray excitation at 300 K & $0.6-2.4$     & \cite{Cherepy:2008,Loef:2009,Glodo:2010,Alekhin:2011} \\

 \hline
 \end{tabular}
  \label{properties}
 \end{center}
 \end{table}

An important advantage of SrI$_2$(Eu) in
comparison to other high resolution scintillators, like for
instance LaCl$_3$(Ce), LaBr$_3$(Ce), Lu$_2$SiO$_5$(Ce), LuI$_3$(Ce),
is the absence of natural long-living radioactive isotopes (as $^{138}$La
in lanthanum and $^{176}$Lu in lutetium). It
makes SrI$_2$(Eu) scintillators promising in various applications, in
particular for low counting experiments as e.g. those searching for double
$\beta$ decay.

The main aim of our study was to test the internal radioactive contamination
of a SrI$_2$(Eu) crystal scintillator. We have also estimated
the response of the detector to $\alpha$ particles by using the data of
low background measurements where events of $^{214}$Po decays
(daughter of $^{226}$Ra from the $^{238}$U chain) were recorded.
As a by-product of the measurements, we have derived limits on
double $\beta$ processes in $^{84}$Sr.

\section{Scintillator, measurements, results and discussion}

\subsection{Development of SrI$_2$(Eu) crystal scintillators}

A single crystal of strontium iodide doped by 1.2\% of Eu\footnote{We present data on the nominal concentration of Eu in the initial powder 
used for the crystal growth. The concentration of Eu in the crystal may be lower due to unknown segregation of Eu in SrI$_2$. Moreover we cannot exclude some nonuniformity of Eu distribution in the crystal volume and, therefore, presence of concentration gradient of Eu.}
was grown in a quartz ampoule using the vertical
Stockbarger method \cite{Stocke}. Anhydrous strontium iodide activated by
europium was obtained by the reaction of the strontium carbonate and
europium oxide with the hydroiodic acid as described in
\cite{Hofstadter:1968}. After drying, the obtained hydrate was
placed in the quartz ampoule for the crystal growth and slowly heated
for five days up to $150^\circ$~C with permanent vacuum pumping.
As a next step the ampoule was welded and placed in the
Stockbarger growing set-up, the temperature was increased up to
$538^\circ$~C. The crystal was grown with a speed of 20 mm per day as
described in \cite{Loef:2009}. The crystal boule was cut in a dry
box filled by pure nitrogen to obtain a near to cylindrical
scintillator 13 mm in diameter and 11 mm length (see Fig. 1, left). The crystal was
wrapped with PTFE tape and encapsulated using epoxy glue in an
oxygen-free high thermal conductivity (OFHC) copper container with
a quartz window, all the materials with low level of radioactive
contamination. It is shown in Fig. 1, right; as one can see, the crystal scintillator is neither milky nor cracked. However, the 
scintillator is not of exact cylindrical shape.

\nopagebreak
\begin{figure}[ht]
\begin{center}
\resizebox{0.42\textwidth}{!}{\includegraphics{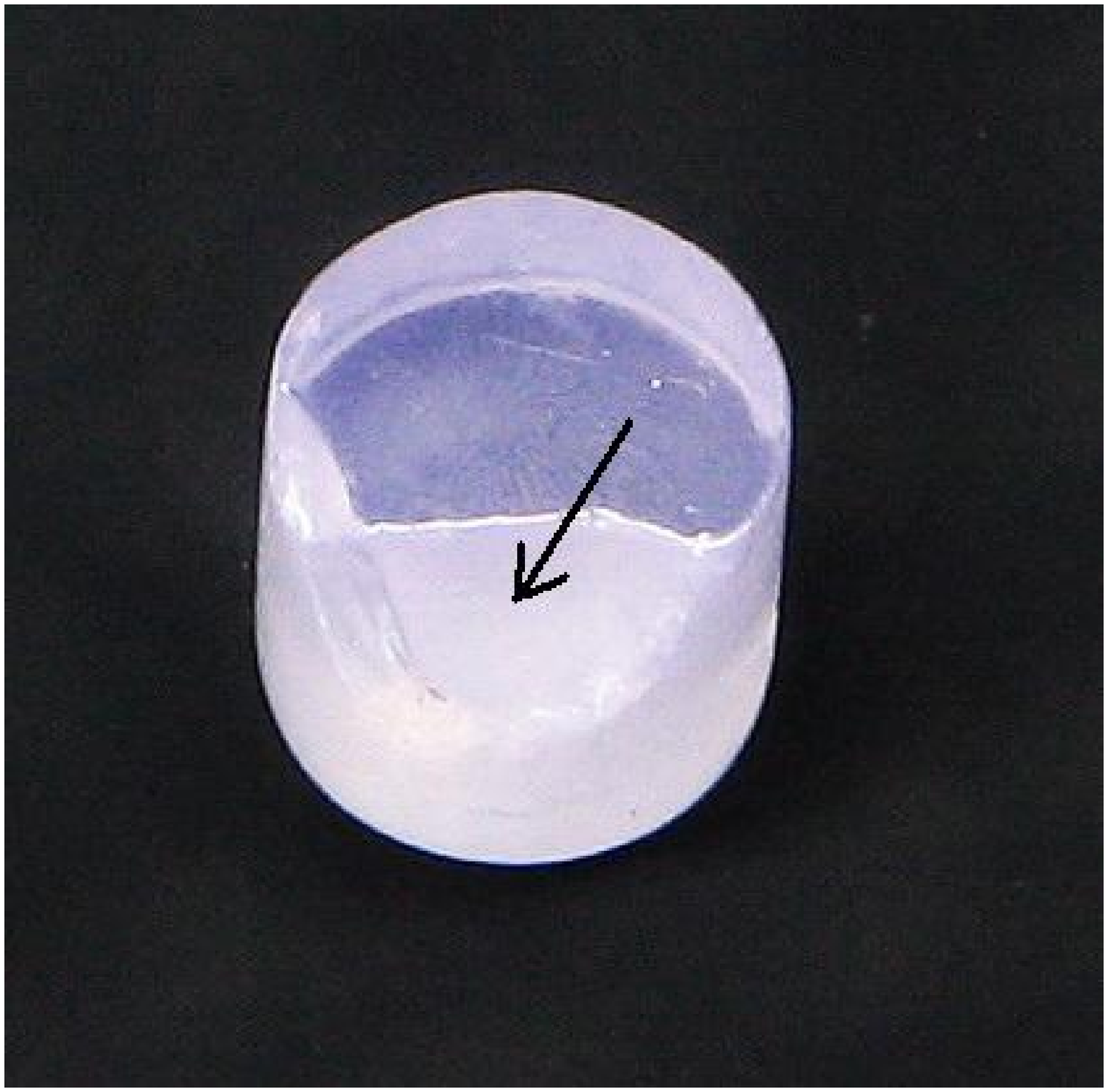}}
\resizebox{0.38\textwidth}{!}{\includegraphics{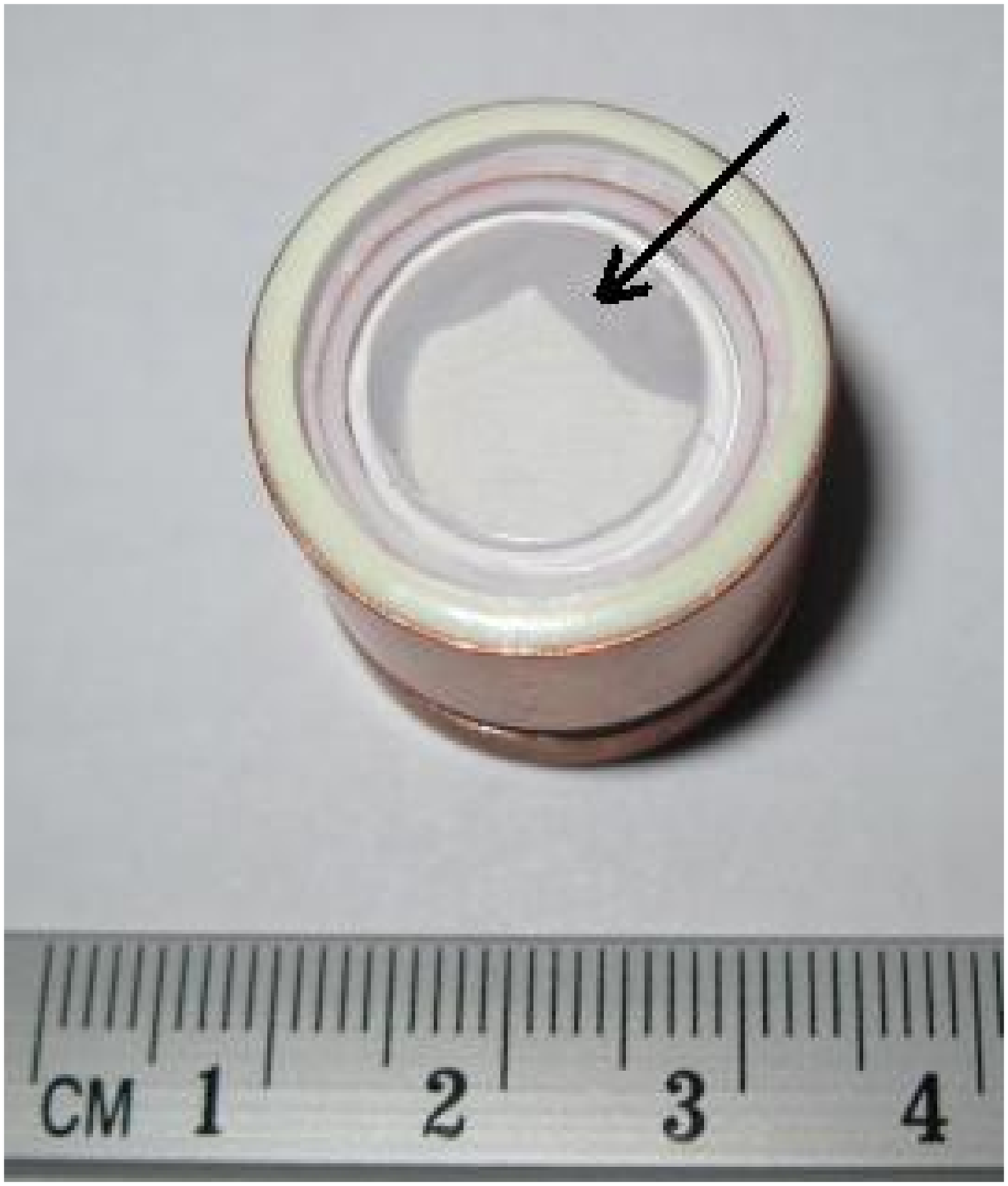}}
 \caption{(Color online) Left: SrI$_2$(Eu) crystal before encapsulation.
Right: Low background SrI$_2$(Eu) scintillation detector. Arrows show irregularity of the crystal shape.}
\end{center}
\end{figure}

\subsection{Energy resolution and relative pulse amplitude}

In order to investigate its scintillation properties, the SrI$_2$(Eu)
crystal scintillator was coupled to a 3" Philips XP2412
photomultiplier (PMT) with a bialkali photocathode using Dow
Corning Q2-3067 optical couplant. The detector was irradiated with
$\gamma$ quanta from $^{60}$Co, $^{137}$Cs, $^{207}$Bi, $^{232}$Th
and $^{241}$Am $\gamma$ sources. The measurements were carried out
using an ORTEC 572 spectrometric amplifier with $10~\mu$s shaping
time and a peak sensitive analog-to-digital converter. Fig. 2
shows the pulse amplitude spectra measured by the SrI$_2$(Eu)
scintillator with $^{60}$Co, $^{137}$Cs, $^{207}$Bi and $^{241}$Am
$\gamma$ sources, respectively. The energy resolution FWHM for the 662 keV $\gamma$
line of $^{137}$Cs is 5.8\%; it is 
worse than the best reported results
(FWHM $=2.6\%-3.7\%$ at 662 keV)
\cite{Cherepy:2008,Loef:2009,Glodo:2010,Cherepy:2009b}. This fact is probably
due to not enough high level of the initial purity of the used powder, to not perfect technology of the crystal production (which is
under development now), to a lower amount of the Eu dopant and to a possible concentration gradient of the Eu in the crystal.
Besides, some degradation of the energy
resolution can be due to the irregular shape of the
crystal (a clear effect of the shape of the SrI$_2$(Eu) scintillators on the energy
resolution is reported in \cite{Sturm:2011}). Moreover, we have used a bialkali PMT, 
while PMTs with a super-bialkali photocathodes have been applied in the works \cite{Cherepy:2008,Loef:2009,Glodo:2010,Sturm:2011,Cherepy:2009b}.

\begin{figure}[t]
\begin{center}
\mbox{\epsfig{figure=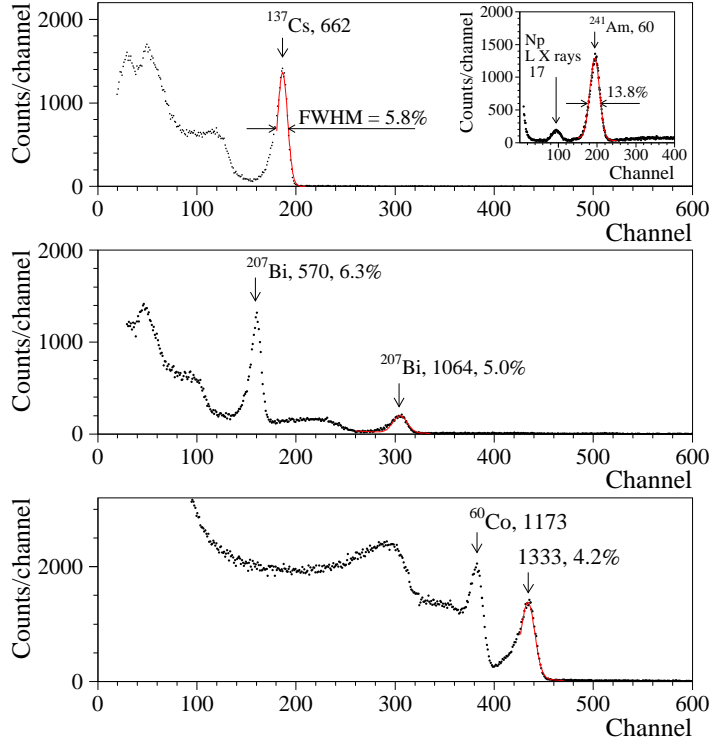,height=10.0cm}} \caption{(Color
online) Energy spectra of $^{137}$Cs, $^{241}$Am (inset),
$^{207}$Bi, and $^{60}$Co $\gamma$ rays measured with the
SrI$_2$(Eu) scintillation crystal. Energies of the $\gamma$ lines
are in keV.}
\end{center}
\end{figure}

The energy resolution of the SrI$_2$(Eu) crystal scintillator
measured in the ($60-2615$) keV energy range is presented in Fig.
3. According to \cite{Dorenbos:1995,Moszynski:2003} the data were fitted (by the chi-square method;
$\chi^2/n.d.f.=9.5/5=1.9$, where n.d.f. is number of
degrees of freedom) by the function FWHM$(\%)=\sqrt{a+b/E_{\gamma}}$
(where $E_{\gamma}$ is the energy of the $\gamma$ quanta
in keV) with parameters $a=(10\pm2)$ and $b=(14200\pm1500)$ keV.
The relative pulse amplitude of the SrI$_2$(Eu) detector was found
to be 87\% of a commercial NaI(Tl) scintillator
($\oslash40$~mm~$\times~40$ mm) (see Fig. 4).

\begin{figure}[ht]
\begin{center}
\mbox{\epsfig{figure=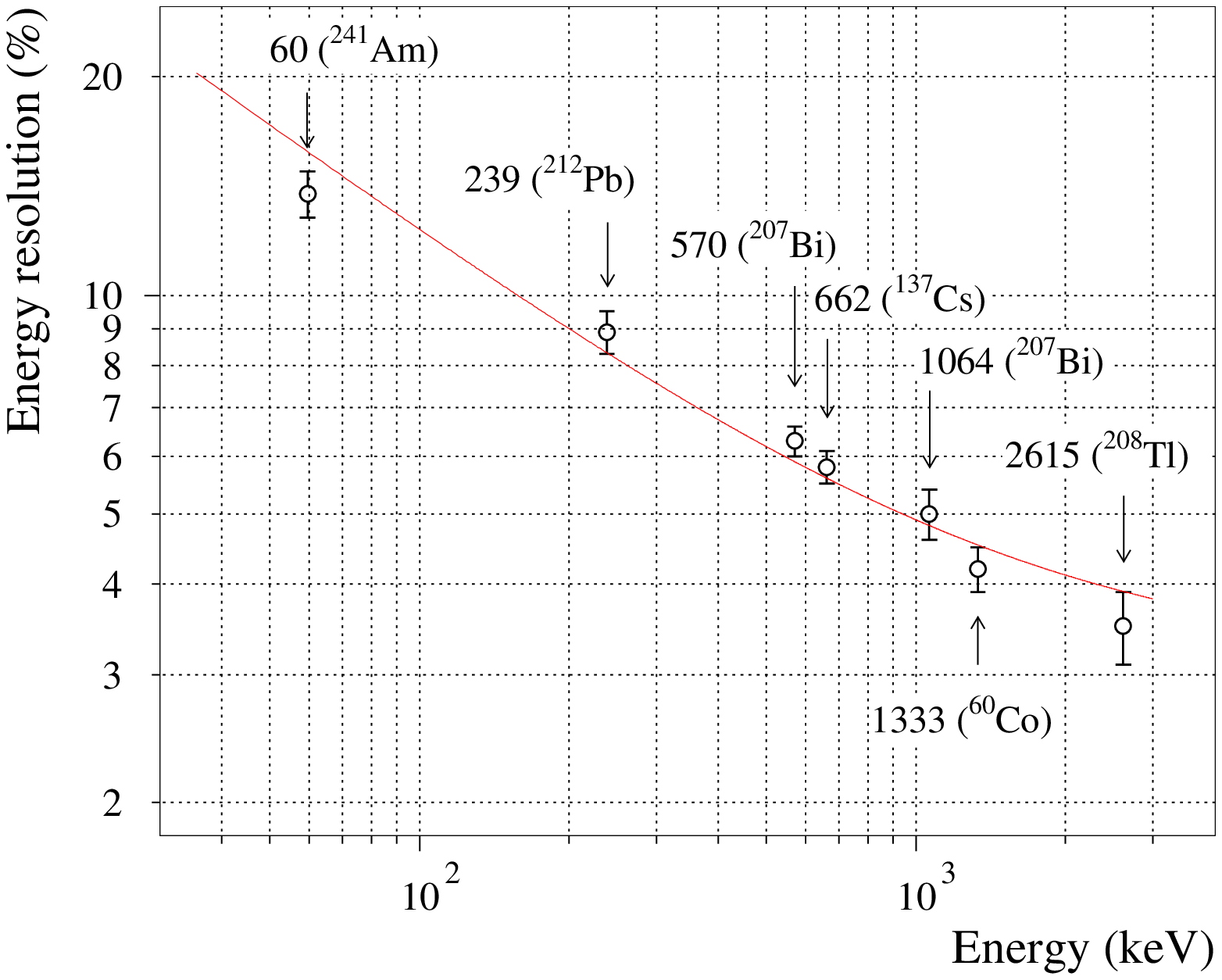,height=9.0cm}} \caption{Energy
resolution (FWHM) versus $\gamma$ energy, measured by the
SrI$_2$(Eu) crystal scintillator.}
\end{center}
\end{figure}

\begin{figure}[ht]
\begin{center}
\mbox{\epsfig{figure=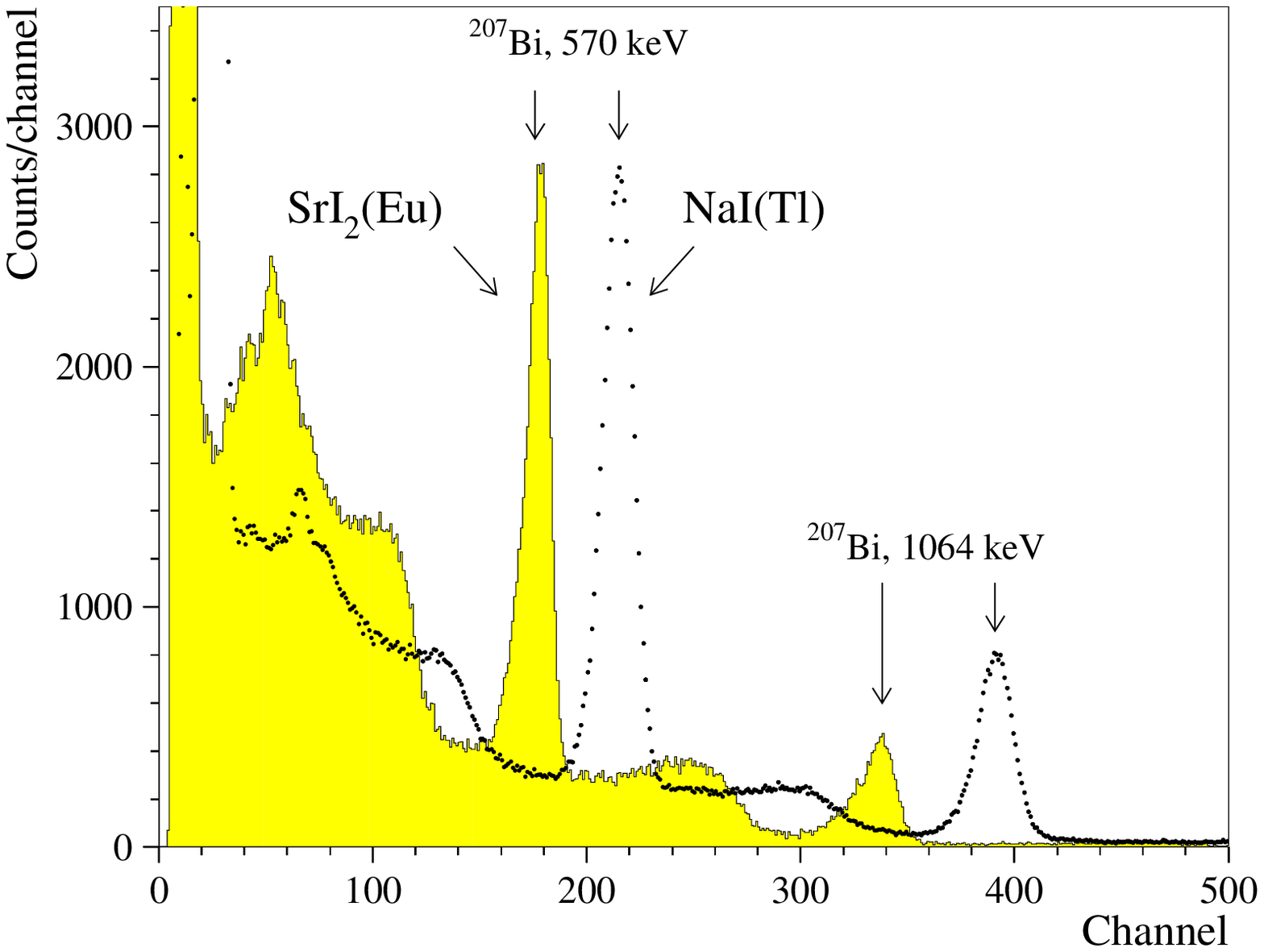,height=8.0cm}}
\vspace{0.2cm}
\caption{Energy
spectra of $^{207}$Bi $\gamma$ rays measured by the SrI$_2$(Eu) and
by a commercial NaI(Tl) scintillation detectors in the low background set up at sea level.}
\end{center}
\end{figure}

It is rather
difficult to derive a photon yield value from the measurements;
indeed, one should know both the light collection and the PMT
quantum efficiency for the scintillation detectors. It should be
stressed that the calculations of the light collection in scintillation
detectors is a rather complicated problem. In addition, we do not know an
emission spectrum of our sample, while in literature there are
different data on this \cite{Cherepy:2008,Loef:2009,Alekhin:2011}.
Nevertheless, taking into account: i) the comparable emission spectra of
SrI$_2$(Eu) and of NaI(Tl) (maximum at $429-436$ nm and at 415 nm, respectively);
ii) the relatively flat behaviour of the PMT spectral sensitivity in
the region $(400-440)$ nm; iii) the typical light
yield of NaI(Tl) in literature: $\approx40\times10^3$ photons/MeV, we can
conclude that the light yield of the sample under study is still
far from the best reported $(68-120)\times 10^3$ photons/MeV
\cite{Cherepy:2008,Loef:2009,Glodo:2010}.

\subsection{Low background measurements in scintillation mode at sea level}

The radioactive contamination of the crystal was measured in the
low background set-up installed at sea-level
in the Institute for Nuclear Research (INR, Kyiv, Ukraine). In the set-up, a SrI$_2$(Eu)
crystal scintillator was optically connected to a 3"
photomultiplier tube Philips XP2412 through a high purity
polystyrene light-guide ($\oslash 66\times120$ mm). The optical
contact between the scintillation crystal, the light-guide and
the PMT was provided by Dow Corning Q2-3067 optical couplant. The
light-guide was wrapped with aluminised Mylar. The detector was
surrounded by a passive shield made of OFHC copper (5-12 cm
thick), and lead (5 cm thick). After the first run of measurements over 52 h,
an anti-muon veto counter was installed above the set-up. The counter consists of polystyrene 
based plastic scintillator $50\times50\times8$ cm 
viewed by a low background PMT FEU-125 (Ekran Optical Systems, Russia) with a diameter 
of the photocathode equal to 15 cm. The anti-muon shield suppressed the background 
caused by cosmic rays by a factor $\approx3$ (at the energy 
$\approx 4$ MeV).

An event-by-event data acquisition system has recorded the pulse shape
of the SrI$_2$(Eu) scintillator over a time window of 100 $\mu$s
(by using a 20 MS/s 12 bit transient digitizer
\cite{Fazzini:1998}), the arrival time of the signals (with an
accuracy of 0.3 $\mu$s), and the signals amplitude by a peak sensitive
analog-to-digital converter.

The energy scale and the energy resolution of the detector were
determined in calibration runs by $^{60}$Co, $^{137}$Cs and $^{207}$Bi
$\gamma$ ray sources. The energy resolution becomes slightly worse
due to the light-guide used in the low background set-up. It can
be fitted by a function: FWHM(\%)$~=\sqrt{a+b/E_{\gamma}}$ with parameteres 
$a=(7.4\pm4.2)$ and $b=(28100\pm8000)$ keV, where $E_{\gamma}$ is the 
energy of the $\gamma$ quanta in keV.

A search for the fast chain $^{214}$Bi ($Q_{\beta}=3272$ keV,
$T_{1/2}=19.9$ m) $\rightarrow $ $^{214}$Po
($Q_{\alpha}=7833$~keV, $T_{1/2}=164~\mu$s) $\rightarrow $
$^{210}$Pb of the $^{238}$U family was performed by analysing
the double pulses (see the technique of the double pulse analysis
e.g. in \cite{Danevich:2003,Belli:2007}); the result of the analysis
is presented in Fig. 5. The obtained energy spectra of the first and
second events, as well as the time distribution between the signals
can be explained by the fast $^{214}$Bi~--~$^{214}$Po decay sequence.
Taking into account the detection efficiency in the time
window $(15-77)~\mu$s (it contains 21.6\% of $^{214}$Po decays),
the mass of the crystal 6.6 g, the measuring time 101.52 h
and the number of selected events (52), one can estimate the activity of
$^{226}$Ra in the SrI$_2$(Eu) crystal as 100(14) mBq/kg.

\nopagebreak
\begin{figure}[t]
\begin{center}
\mbox{\epsfig{figure=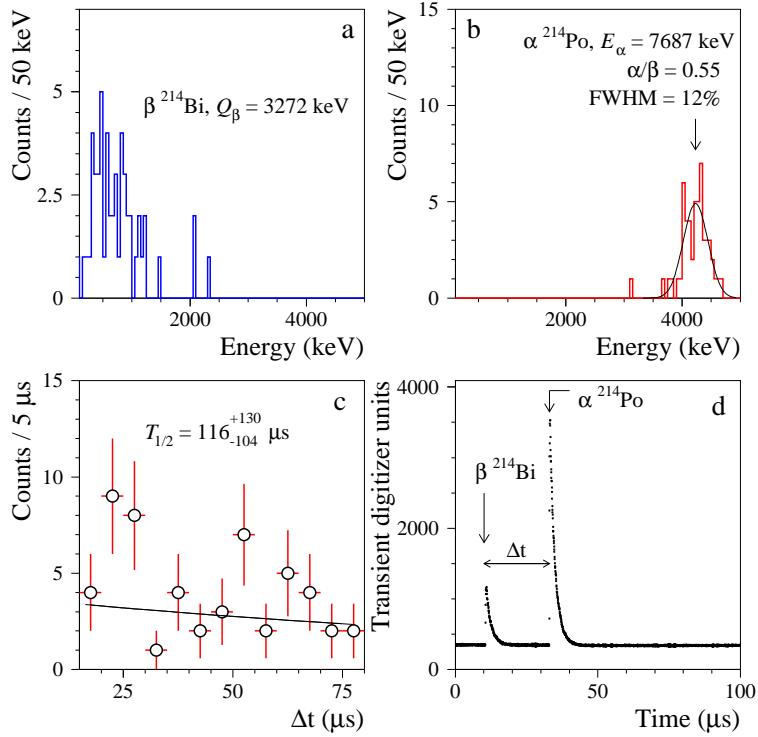,height=10.0cm}} \caption{(Color
online) The energy (a) and (b), and time (c) distributions for the
fast sequence of $\beta$ ($^{214}$Bi, $Q_{\beta}=3272$ keV) and
$\alpha$ ($^{212}$Po, $E_{\alpha}=7687$ keV, $T_{1/2}=164.3(20)
~\mu$s \cite{ToI98}) decays  by the analysis of double pulses in
the background data accumulated over 101.52 h. (d) Example of such an
event in the SrI$_2$(Eu) scintillator.}
\end{center}
\end{figure}

By using the result of the double pulses analysis, we have
estimated the response of the SrI$_2$(Eu) crystal scintillator to $\alpha$
particles. The quenching of the scintillation light yield can be expressed
through the so called $\alpha/\beta$ ratio, which is
the ratio of the position of an $\alpha$ peak in the energy scale 
measured with $\gamma$ quanta to the energy of the $\alpha$
particles. Considering the spectrum presented in Fig. 5(b) as given 
by $\alpha$ particles of $^{214}$Po with energy 7687 keV, we
can estimate the $\alpha/\beta$ ratio as 0.55. A similar
quenching was observed in ref. \cite{Sysoeva:1998}
for NaI(Tl) ($\alpha/\beta=0.66$) and
CsI(Tl) ($\alpha/\beta=0.67$) crystal scintillators with the $\alpha$
particles of a $^{241}$Am source with energy 5.48 MeV;
in ref. \cite{DAMA} similar values have been
measured in ultra low background NaI(Tl) for $\alpha$ trace contaminants
internal to the crystals. It should be stressed that the energy
resolution
for the $\alpha$ peak of $^{214}$Po (FWHM$_{\alpha}~=12\%$) is
worse than that expected at the energy $\approx 4.2$ MeV according to the calibration
with $\gamma$ sources (FWHM$_{\gamma}~\approx3\%$). A similar
effect was observed in crystal scintillators with anisotropic
crystal structure, as for instance in CdWO$_4$
\cite{Danevich:2003} and ZnWO$_4$ \cite{Belli:2009}.

A search for internal contamination of the crystal by $^{228}$Th
(daughter of $^{232}$Th) was realized with the help of the
time-amplitude analysis\footnote{The method of the time-amplitude
analysis is described in detail in
\cite{Danevich:1995,Danevich:2001}.}. To determine the activity of
$^{228}$Th, the following sequence of $\alpha $ decays was
selected: $^{220}$Rn ($Q_\alpha $ = $6405$ keV, $T_{1/2}$ = $55.6$
s) $ \to $ $^{216}$Po ($Q_\alpha $ = $6907$ MeV, $T_{1/2}$ =
$0.145$ s) $\to $ $^{212}$Pb. Assuming that the $\alpha/\beta$
ratio for the $\alpha$ particles of ($6.3-7.7$) MeV is in the range
of $0.5-0.6$, the energy interval for both $\alpha$ particles was
chosen as ($2.8-4.5$) MeV. The result of the selection is presented in
Fig. 6.

\begin{figure}[htbp]
\begin{center}
\mbox{\epsfig{figure=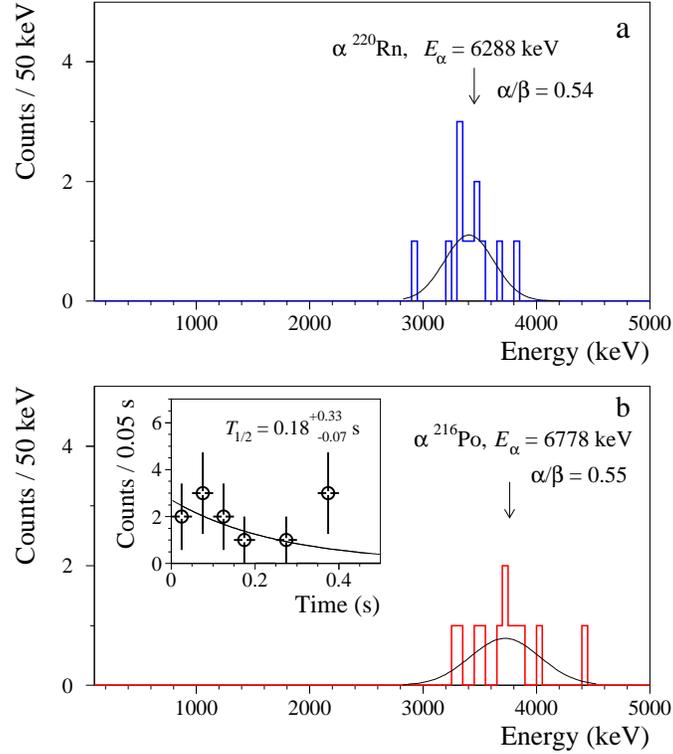,height=10.0cm}}
\caption{(Color online) Energy spectra of the first (a) and second (b) events
selected by the time-amplitude analysis (see text) from the background
data accumulated over 101.52 h with the SrI$_2$(Eu) detector. In the
inset the time distribution between the first and second events
together with an exponential fit are presented. The positions of the
selected events (solid lines represents fit of the data by Gaussian functions)
and the obtained
half-life of $^{216}$Po ($0.18^{+0.33}_{-0.07}$ s, the table value
is 0.145(2) s \cite{ToI98}) do not contradict the assumption that
these events are caused by the sequence of $\alpha$ decays
$^{220}$Rn$\to^{216}$Po$\to^{212}$Pb.}
\end{center}
\end{figure}

Despite the low statistics (only 12 pairs were found), the
positions of the selected events and the distribution of the time
intervals between the events do not contradict the expectations for the
$\alpha $ particles of the chain. Taking into account the
efficiency in the time window $(0.01-0.5)$ s to select $^{216}$Po
$\to $ $^{212}$Pb events (86.2\%), the activity of $^{228}$Th in
the crystal can be calculated as 6(2) mBq/kg.

The energy spectrum measured with the SrI$_2$(Eu) scintillator
over 101.52 h is presented in Fig. 7.
There is a peak in the spectrum at the energy of $(665\pm 5)$ keV,
which can be explained by contaminations of the SrI$_2$(Eu)
detector or/and of the set-up by $^{137}$Cs (probably as a result
of pollution after the Chernobyl accident). Taking into account that
our set-up is installed at sea level, a significant part of the
background above 2.6 MeV (the edge of the $\gamma$ quanta energy from the
natural radioactivity) can be attributed to cosmic rays.

\begin{figure}[ht]
\begin{center}
\mbox{\epsfig{figure=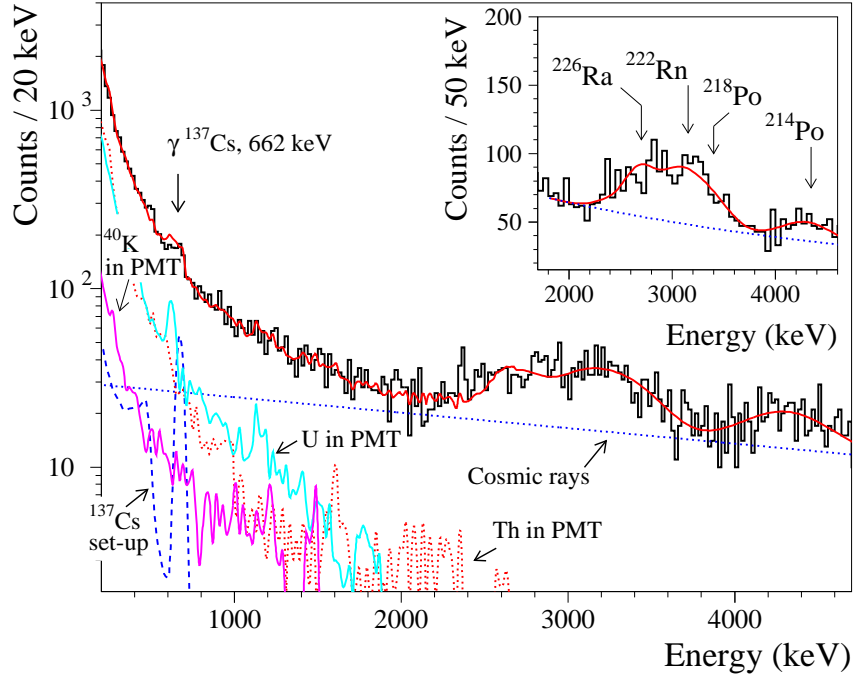,height=9.0cm}} \caption{(Color
online) Energy spectrum of the SrI$_2$(Eu) scintillator measured
over 101.52 hours together with the model
of the background. The main components of the external background ($^{137}$Cs
on details of the set-up, $^{40}$K, U and Th in PMT) and of the cosmic
rays are shown (see text for details). (Inset) Increase of the counting
rate in the energy region ($2.5-4.3$) MeV is due to
$\alpha$ activity of trace U/Th contamination (mainly $^{226}$Ra
with its daughters) of the crystal scintillator.}
\end{center}
\end{figure}

Peculiarities in the spectrum in the energy region $(2.5-4.5)$ MeV
can be explained by the decays of $\alpha$ active U/Th daughters
present in the crystal as trace contamination. To estimate the
activity of the $\alpha$ active nuclides from the U/Th families in
the crystal, the energy spectrum was fitted in the energy interval
$(1.8-4.7)$ MeV by using fourteen Gaussian functions to describe
$\alpha$ peaks of $^{232}$Th, $^{228}$Th (and daughters:
$^{224}$Ra, $^{220}$Rn, $^{216}$Po, $^{212}$Bi), $^{238}$U (and
daughters: $^{234}$U, $^{230}$Th), $^{226}$Ra with daughters
($^{222}$Rn, $^{218}$Po, $^{214}$Po, $^{210}$Po)\footnote{We
assume a broken equilibrium of the U/Th chains in the
scintillator.} plus an exponential function to describe
background\footnote{At sea level the energy spectrum of cosmic ray induced
background in different nuclear detectors, including
scintillating, has a monotonic character, which can be
approximated in reasonable narrow energy regions (a few MeV) by an
exponential function
\cite{Reeves:1984,Kamikubota:1986,Arpesella:1996,Iwawaki:1998}.}.
A fit of the spectrum is shown in the Inset of Fig. 7. The main
contribution to the $\alpha$ activity in the scintillator gives
an activity $(91\pm8)$ mBq/kg for the 
$^{226}$Ra and daughters ($^{222}$Rn, $^{218}$Po and $^{214}$Po). 
This estimate is in
agreement with the result of the double pulse analysis ($100\pm14$;
mBq/kg, see above) and with the measurements performed deep underground
with ultra-low background HPGe $\gamma$
ray spectrometry (Section 2.5). Because of the low statistics and of the relatively poor energy
resolution for $\alpha$ particles, we conservatively give limits
on activities of $^{232}$Th, $^{238}$U and $^{210}$Po in the
SrI$_2$(Eu) scintillator. The data obtained from the fit are
presented in Table  \ref{rad-cont}.

To estimate the contamination of the scintillator by $^{40}$K,
$^{60}$Co, $^{90}$Sr$-^{90}$Y, $^{137}$Cs, $^{138}$La, $^{152}$Eu,
$^{154}$Eu, $^{176}$Lu, $^{210}$Pb$-^{210}$Bi,
their decays in the SrI$_2$(Eu) detector were simulated
with the GEANT4 package \cite{GEANT4} and the event generator
DECAY0 \cite{DECAY4}. The radioactive contamination of the set-up, in
particular, the radioactivity of the PMT can contribute to the
background, too. Therefore we have also simulated the contribution from
the contamination of the PMT by $^{232}$Th, $^{238}$U (with
their daughters) and $^{40}$K. An exponential function was adopted
to describe the contribution of the cosmic rays in the sea level installation
where these measurements were carried out.

Apart from the peak of $^{137}$Cs and the $\alpha$ peaks in the energy region $(2.5-4.3)$ MeV, there are no
other peculiarities in the spectrum which could be ascribed to internal trace
radioactivity. However, even in the case of $^{137}$Cs we cannot
surely distinguish the contribution of internal and external
contamination. Therefore, only limits on contaminations of the
crystal by the possible radionuclides were set on the basis of the
experimental data. With this aim the spectrum was fitted in the
energy interval $(0.2-4.65)$ MeV by the model composed by the
background components ($^{40}$K, U and Th in PMT, pollution of the
set-up surface by $^{137}$Cs, the $\alpha$ peaks of $^{226}$Ra
with daughters, and cosmic rays) plus a distribution of possible
internal radioactive contamination to be estimated. The result of the
fit is presented in Fig. 7 together with the main components of the
background.

The summary on activities (or limits) obtained by the analysis of
the experimental data accumulated at the sea level low background
scintillation set-up is presented in Table \ref{rad-cont}.

Despite the sea level location and the modest shield of the scintillation set-up, 
the measurements allowed the detection of the internal contamination of the 
scintillator by $^{226}$Ra and $^{228}$Th; besides, we have estimated limits 
on activities of $^{238}$U, $^{232}$Th, $^{210}$Pb, $^{210}$Po, 
$^{90}$Sr. It should be stressed that these radionuclides are rather hard to analyse 
with the help of low background HPGe $\gamma$ spectrometry due to 
the absence of noticeable $\gamma$ rays. Moreover, the measurements allowed us
to estimate the $\alpha/\beta$ ratio and to measure the pulse shape for 
$\alpha$ particles (see the next Section) by recording the pulse profiles 
of $^{214}$Po $\alpha$ events inside the scintillator.

\subsection{Pulse shape of scintillation for $\gamma$ quanta ($\beta$
particles) and $\alpha$ particles from the sea level measurements}

The pulse profiles of 41 $\alpha$ events of $^{214}$Po were selected
with the help of the double pulse analysis from the data of the
low background measurements (see Section 2.3 and Fig. 5). The sum
of the pulses is presented in Fig. 8 where also the sum of
approximately two thousands of background $\gamma$ ($\beta$)
events with energies $\approx1.5$ MeV is drawn.

\begin{figure}[ht]
\begin{center}
\mbox{\epsfig{figure=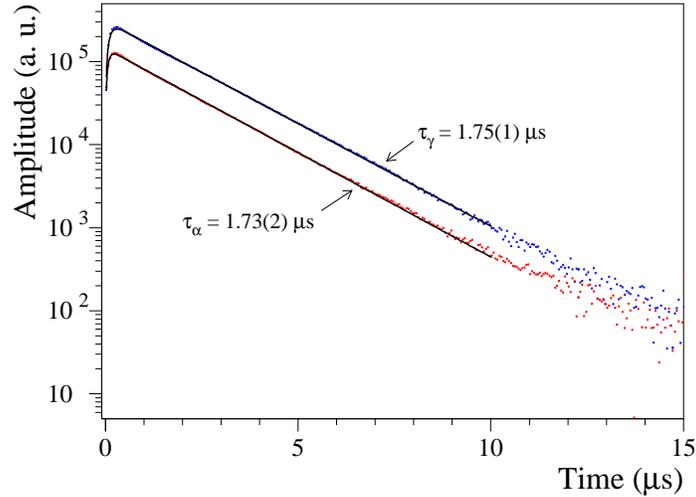,height=7.0cm}} \caption{(Color
online) Scintillation pulse profiles in the SrI$_2$(Eu) crystal measured
for $\gamma$ and $\alpha$ excitation. Fitting functions for
$\gamma$ and $\alpha$ pulses are shown by solid lines.}
\end{center}
\end{figure}

The distributions
were fitted in the time interval $(0-10)~\mu$s by the following
function:

\begin{center}
$f(t)=A(e^{-t/\tau}-e^{-t/\tau _{0}})/(\tau-\tau_{0}),\qquad t>0$,
\end{center}
where $A$ is the intensity (in arbitrary units), and $\tau$ is the decay
constant of the light emission; $\tau_{0}$ is the integration constant of the
electronics ($\approx 0.08~\mu$s). The fit gives the scintillation
decay times in the SrI$_2$(Eu) crystal scintillator:
$\tau_{\gamma}=(1.75\pm0.01)~\mu$s and $\tau_{\alpha}=(1.73\pm0.02)~\mu$s
for $\gamma$ quanta ($\beta$ particles) and $\alpha$ particles,
respectively. Therefore we have not observed any clear indication on
differences in the kinetics of the scintillation decay in the SrI$_2$(Eu)
crystal scintillator under $\gamma$ quanta ($\beta$ particles) and
$\alpha$ particles irradiation.

\subsection{Measurements with ultra-low background HPGe $\gamma$
ray spectrometry deep underground}

The SrI$_2$(Eu) crystal scintillator was measured for 706 h with
the ultra-low background HPGe $\gamma$ ray spectrometer GeCris.
The detector has a volume of 468 cm$^3$ and a 120\% efficiency relatively 
to a 3 in. $\times$ 3 in. NaI(Tl). This detector has a rather thin 
Cu window of 1 mm thickness. The passive shield of the detector consists of 15 cm of OFHC copper and 20 cm of low radioactive lead. The whole 
set-up is sealed in an air-tight plexiglass box continuously flushed with high purity nitrogen gas to avoid the presence of residual environmental radon. The facility is located deep underground in
the Gran Sasso National Laboratories of the I.N.F.N. (average overburden of 3600 m water equivalent) \cite{Arpesella:2002, Laubenstein:2004}. The background
data were accumulated over 1046 h (see Fig. 9).

\begin{figure}[ht]
\begin{center}
\mbox{\epsfig{figure=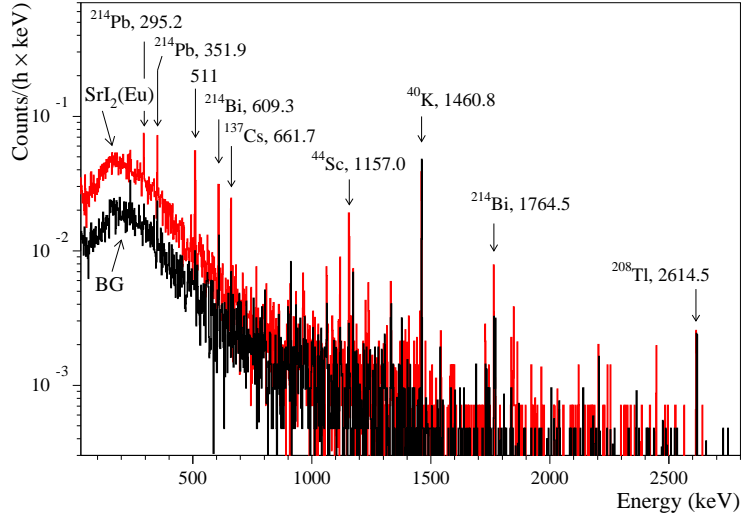,height=7.0cm}} \caption{(Color
online) Energy spectra accumulated with the SrI$_2$(Eu) sample over
706 h, and without sample over 1046 h (BG) by ultra-low background
HPGe $\gamma$ spectrometer deep underground. The energies of the
$\gamma$ lines are in keV.
The background was accumulated before the contamination of the HPGe
set-up by $^{44}$Ti.}
\end{center}
\end{figure}

In order to
determine the radioactive contamination of the sample, the
detection efficiencies were calculated using a Monte Carlo
simulation based on the GEANT4 software package \cite{GEANT4}.
The peaks in the measured spectra are due to the naturally occurring
radionuclides of the uranium and thorium chains, $^{40}$K and
$^{137}$Cs. We have detected contaminations by $^{137}$Cs and
$^{226}$Ra (the $\gamma$ lines of $^{137}$Cs, $^{214}$Bi and
$^{214}$Pb were observed) in the crystal scintillator at the level
of 53(11) mBq/kg and 120(50) mBq/kg, respectively, while limits
were obtained for other potential contaminations. The measured
activities and the limits are presented in Table \ref{rad-cont}.

In addition, we have observed the 1157 keV peak of $^{44}$Sc in the
data accumulated with the SrI$_2$(Eu) crystal
(at a rate of $6.1(9)\times10^{-2}$ counts/h). However, this peak
was due to a contamination of the used HPGe detector
(not of the crystal scintillator sample) by $^{44}$Ti.

\nopagebreak
\begin{table}[ht]
\caption{Radioactive contamination of the SrI$_2$(Eu) scintillator.
The upper limits are given at 90\% C.L., and the uncertainties of
the measured activities at 68\% C.L.}
\begin{center}
\begin{tabular}{llll}
\hline
 Chain      &  Nuclide              & \multicolumn{2}{c}{Activity (mBq/kg)} \\
 \cline{3-4}
 ~          & (Sub-chain)           & Measured in           & Measured by HPGe    \\
 ~          & ~                     & scintillation mode    & ~ \\
 \hline
 ~          & $^{40}$K              & $\leq 200$            & $\leq 255$ \\
 ~          & $^{60}$Co             & $\leq 540$            & $\leq 16$ \\
 ~          & $^{90}$Sr$-^{90}$Y    & $\leq 90$             & ~ \\
 ~          & $^{137}$Cs            & $\leq 140$            & $53\pm11$  \\
 ~          & $^{138}$La            & $\leq 1100$           & $\leq 20$ \\
 ~          & $^{152}$Eu            & $\leq 840$            & $\leq 108$ \\
 ~          & $^{154}$Eu            & $\leq 910$            & $\leq 67$ \\
 ~          & $^{176}$Lu            & $\leq 970$            & $\leq 143$ \\
 $^{232}$Th & $^{232}$Th            & $\leq 3$              & ~ \\
 ~          & $^{228}$Ac            & ~                     & $\leq 68$ \\
 ~          & $^{228}$Th            & $6\pm 2$              & $\leq 52$  \\
 $^{238}$U  & $^{238}$U             & $\leq 40$             & ~ \\
 ~          & $^{226}$Ra            & $100\pm14$            & $120\pm50$ \\
 ~          & $^{210}$Pb$-^{210}$Bi & $\leq 180$            & ~ \\
 ~          & $^{210}$Po            & $\leq 60$             & ~ \\
 \hline
\end{tabular}
\end{center}
\label{rad-cont}
\end{table}

The radioactive purity of the SrI$_2$(Eu) scintillator is
still far from that of NaI(Tl) and CsI(Tl) scintillators (especially
those developed with high radiopurity for dark matter search \cite{DAMA,KIMS}).
At the same time it is much better than
the typical purity
of LaCl$_3$(Ce), LaBr$_3$(Ce), Lu$_2$SiO$_5$(Ce) and
LuI$_3$(Ce) crystal scintillators (see Table \ref{rc-comp}, where the
radioactive contamination of the SrI$_2$(Eu) crystal is compared
with that of NaI(Tl), CsI(Tl) and scintillators containing La or Lu).

\begin{table*}[ht]
\caption{Radioactive contaminations of SrI$_2$(Eu) crystal
scintillator. Data for NaI(Tl), CsI(Tl), LaCl$_3$, LaBr$_3$(Ce),
Lu$_2$SiO$_5$, LuI$_3$(Ce) are given for comparison.
Activities of LaBr$_3$(Ce), Lu$_2$SiO$_5$(Ce) and LuI$_3$(Ce) are calculated values
based on the $^{138}$La and $^{176}$Lu half-lives \cite{ToI98},
on the abundances of the isotopes \cite{Berglund:2011} and on the chemical formula of the compounds.}

\begin{center}
\begin{tabular}{llllll}
\hline
Scintillator                      & \multicolumn{5}{c}{Activity (mBq/kg)} \\
~                                 & $^{40}$K  & $^{138}$La      & $^{176}$Lu      & $^{226}$Ra & $^{228}$Th   \\
\hline
SrI$_2$(Eu)$^a$                   & $\leq255$ & $\leq20$        & $\leq143$       & 120        & $\leq11$     \\
NaI(Tl) \cite{DAMA}               & $< 0.6$   & ~               & ~               & $\sim0.02$ & $\sim 0.009$ \\
CsI(Tl) \cite{KIMS}               & ~         & ~               & ~               & 0.009      & 0.002        \\
LaCl$_3$(Ce) \cite{Bernabei:2005} & ~         & $4.1\times10^5$ & ~               & $\leq 35$  & $\leq0.36$   \\
LaBr$_3$(Ce)                      & ~         & $3.0\times10^5$ & ~               & ~          &              \\
Lu$_2$SiO$_5$(Ce)                 & ~         & ~               & $3.9\times10^7$ & ~          & ~            \\
LuI$_3$(Ce)                       & ~         & ~               & $1.6\times10^7$ & ~          & ~            \\
\hline
$^a$~This work
\end{tabular}
\end{center}
\label{rc-comp}
\end{table*}

\section{Search for 2$\beta$ decay of $^{84}$Sr}

The data of the low background measurements with the HPGe detector can
be used to search for double $\beta$ processes in $^{84}$Sr
accompanied by the emission of $\gamma$ quanta. The decay scheme of
$^{84}$Sr is presented in Fig. 10. The energy of double $\beta$
decay of $^{84}$Sr is comparatively high: $Q_{2\beta}=1787(4)$ keV
\cite{Audi:2003}, however the isotopic abundance is rather low:
$\delta = 0.56(1)\%$ \cite{Berglund:2011}.

We do not observe any peaks in the spectrum accumulated with the
sample of the SrI$_2$(Eu) scintillator which could indicate double
$\beta$ activity of $^{84}$Sr. Therefore, only lower half-life
limits ($\lim T_{1/2}$) can be set according to the formula: $\lim
T_{1/2} = N \cdot \eta \cdot t \cdot \ln 2 / \lim S$, where $N$ is
the number of $^{84}$Sr nuclei in the sample, $\eta$ is the
detection efficiency, $t$ is the measuring time, and $\lim S$ is
the number of events of the effect searched for which can be
excluded at given confidence level (C.L.; all the limits obtained
in the present study are given at 90\% C.L.). The efficiencies of
the detector for the double $\beta$ processes in $^{84}$Sr were
calculated with the GEANT4 code \cite{GEANT4} and DECAY0 event
generator \cite{DECAY4}.

\begin{figure}[htb]
\begin{center}
 \mbox{\epsfig{figure=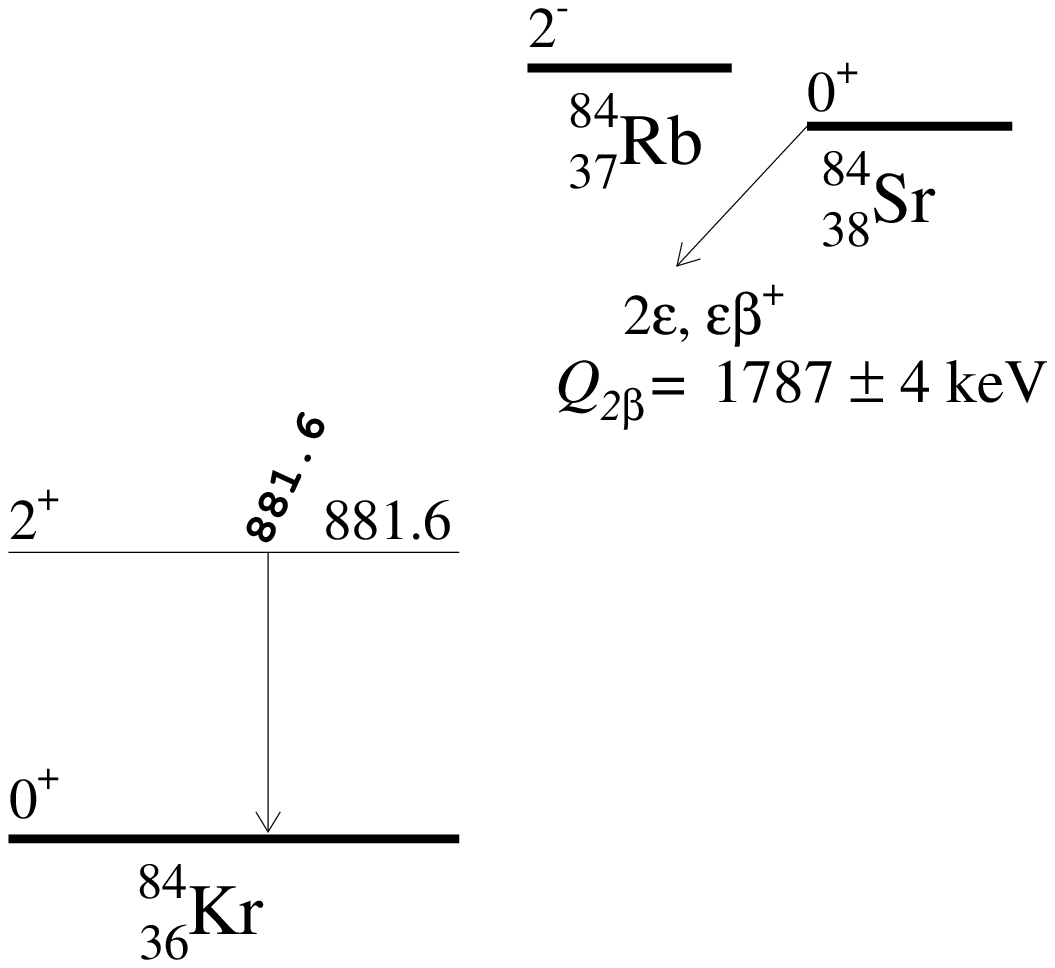,height=5.5cm}}
 \caption{Expected decay scheme of $^{84}$Sr \cite{ToI98}. The energies of the
excited levels and of the emitted
$\gamma$ quanta are in keV.}
\end{center}
\end{figure}

One positron can be emitted in the $\varepsilon\beta^+$ decay of
$^{84}$Sr with energy up to $(765\pm4)$ keV. The annihilation of
the positron will give rise to two 511 keV $\gamma$'s leading to an extra
rate in the annihilation peak. The part of the spectrum in the energy
interval $(450-550)$ keV is shown in Fig. 11.

\begin{figure}[ht]
\begin{center}
\mbox{\epsfig{figure=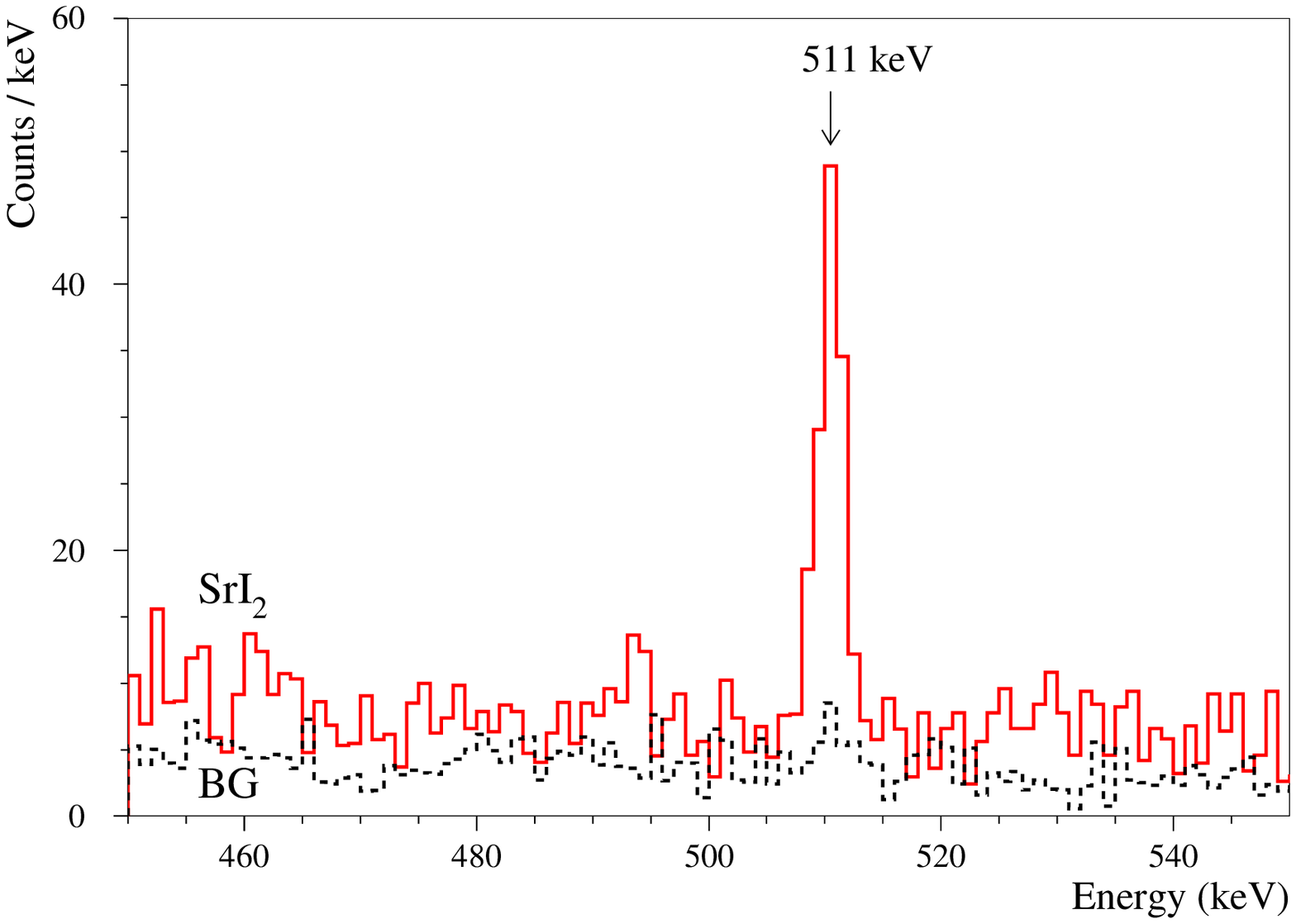,height=6.5cm}} \caption{(Color
online) Energy spectra accumulated with the SrI$_2$(Eu) sample over
706 h, and without sample over 1046 h (BG) by ultra-low background
HPGe $\gamma$ spectrometer deep underground. The spectra are normalized on the time
of the measurements with the SrI$_2$(Eu) sample.}
\end{center}
\end{figure}

There are
peculiarities in both the spectra accumulated with the SrI$_2$(Eu)
sample [$(111\pm14)$ counts at $(510.9\pm0.2)$~keV] and in the
background [($12\pm5)$ counts at $(510.8\pm0.3)$~keV], which
can be ascribed to annihilation peaks. The main contribution to
the 511 keV peak [($108 \pm 22)$ counts] is coming from decays of
$^{44}$Sc (daughter of $^{44}$Ti) present in the HPGe detector
as contamination (see Section 3.4), $(8\pm3)$ counts corresponds to
the background of the detector before the contamination. The
difference in the areas of the annihilation peak: ($-5\pm26$)
counts, which can be attributed to electron capture with positron
emission in $^{84}$Sr, gives no indication on the effect. In
accordance with the Feldman-Cousins procedure \cite{Feldman:1998}
(here and hereafter we use this approach to estimate the values of $\lim
S$ for all the processes searched for) we should take $\lim S=38$
counts which can be excluded at 90\% C.L. Taking into account the
number of $^{84}$Sr nuclei in the sample ($6.5\times10^{19}$) and
the detection efficiency ($\eta=7.2\%$), we have calculated the
following limit on the half-life of $^{84}$Sr relatively to
$\varepsilon\beta^+$ decay:

\begin{center}
$T_{1/2}^{(2\nu+0\nu)\varepsilon\beta^{+}}($g.s.$~\rightarrow~$g.s.$)\geq
6.9\times10^{15}$ yr.
\end{center}

We cannot study the $2\nu2K$ capture in $^{84}$Sr to the ground
state of $^{84}$Kr because the energies of the expected X rays
after the decay are too low in energy
(the binding energy of electrons at $K$ shell of
krypton atom is only 14.3 keV \cite{ToI98} while the energy
threshold of the HPGe detector is $\approx20$ keV).

In the neutrinoless double electron capture to the ground state of
the daughter nucleus, in addition to the X rays, some other
particle(s) must be emitted to take away the rest of the energy.
Usually one bremsstrahlung $\gamma$ quantum is assumed. The energy
of the $\gamma$ quantum is expected to be equal to
$E_\gamma=Q_{2\beta}-E_{b1}-E_{b2}$, where $E_{b1}$ and $E_{b2}$
are the binding energies of the first and of the second captured
electrons on the atomic shell. The binding energies on the $K$,
$L_1, L_2$ and $L_3$ shells in Kr are equal to $E_K=14.3$ keV,
$E_{L_1}=1.9$ keV, $E_{L_2}\approx E_{L_3}=1.7$ keV, respectively \cite{ToI98}.
Therefore, the expected energies of the $\gamma$ quanta for the
$0\nu2\varepsilon$ capture in $^{84}$Sr to the ground state of
$^{84}$Kr are in the intervals: i) $E_\gamma=(1754-1762)$ keV for
the $0\nu 2K$; ii) $E_\gamma=(1767-1775)$ keV for the $0\nu KL$;
iii) $E_\gamma=(1779-1788)$ keV for the $0\nu 2L$ process.

No events are detected (see Fig. 12, a) in the energy intervals $(1754-1762)$ and
$(1779-1788)$ keV, where the g.s. $\rightarrow$ g.s. $0\nu2K$
and $0\nu2L$ decay of $^{84}$Sr is expected.

\begin{figure}[htbp]
\begin{center}
\mbox{\epsfig{figure=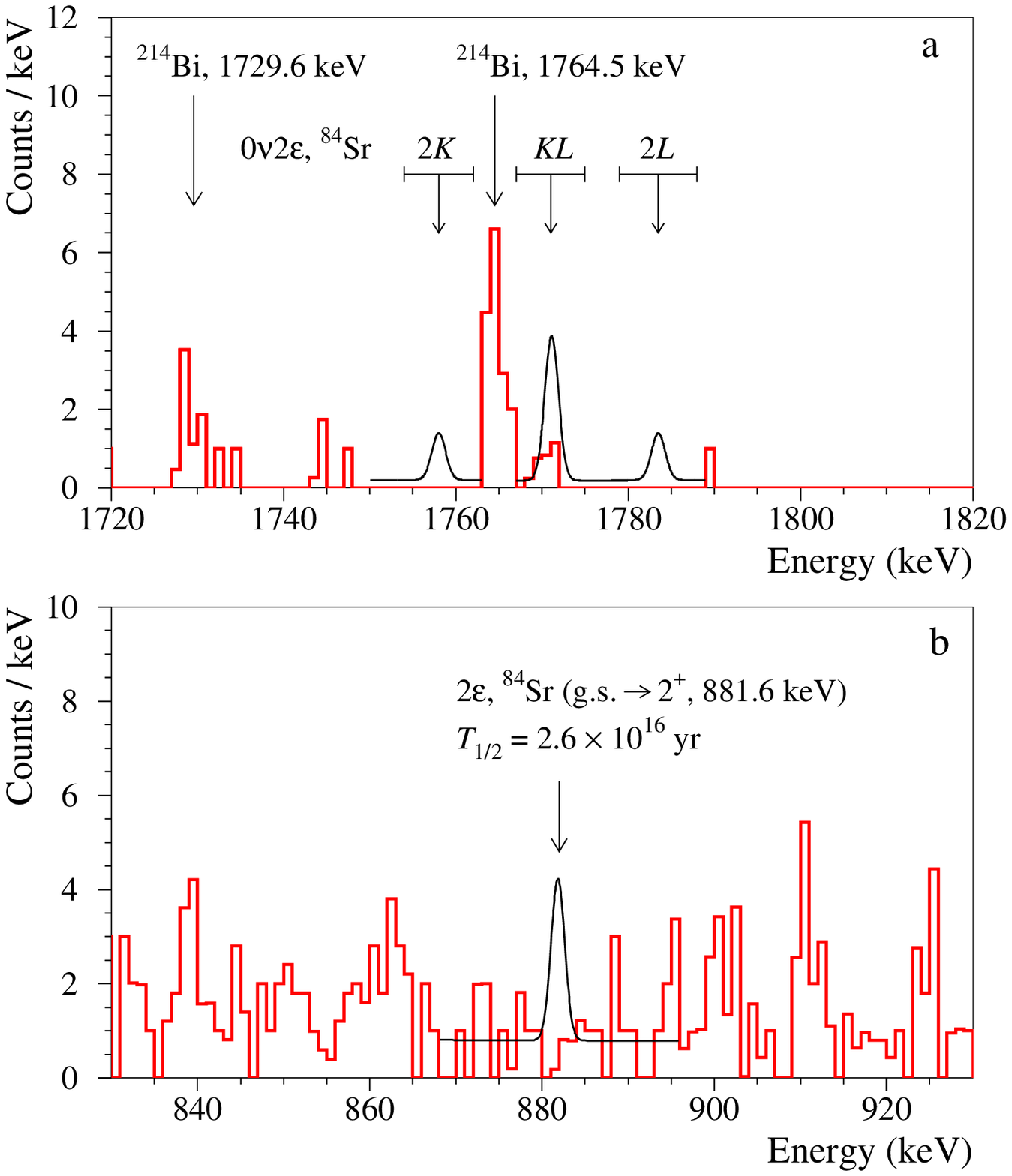,height=10.0cm}} \caption{(Color
online) (a) Part of the energy spectrum measured with the
SrI$_2$(Eu) sample in the energy region where peaks from the $0\nu2\varepsilon$
processes in $^{84}$Sr ($2K$, $KL$ and $2L$) to the ground state of
$^{84}$Kr are expected. The excluded peaks for the processes are
shown by solid lines. (b) Part of the spectrum in the energy
interval $(830-930)$ keV where a $\gamma$ peak with the energy of
881.6 keV is expected for the $2\varepsilon$ decay of $^{84}$Sr to the
excited level $2^+$ of $^{84}$Kr. The area of the peak, shown by solid
line, corresponds to the half-life $2.6\times10^{16}$ yr excluded at
90\% C.L.}
\end{center}
\end{figure}

According to
\cite{Feldman:1998} we should take 2.4 events as $\lim S$.
Therefore, taking into account the detection efficiencies of the
effects (4.0\% and 3.9\%, respectively), we can set the following
limits on the processes:

 \begin{center}
 $T_{1/2}^{0\nu 2K}$(g.s.$~\rightarrow~$g.s.$)\geq~6.0\times10^{16}$ yr,

 $T_{1/2}^{0\nu 2L}$(g.s.$~\rightarrow~$g.s.$)\geq~5.9\times10^{16}$ yr.
 \end{center}

There are 3 events at energy $\approx1770$ keV (due to $^{207}$Bi in
the background) where the $0\nu
KL$ decay of $^{84}$Sr is expected. Taking in this case $\lim S=7.4$
counts, while the detection efficiency is 3.9\%, one can
obtain the following half-life limit on the $0\nu KL$ process in
$^{84}$Sr:

 \begin{center}
 $T_{1/2}^{0\nu KL}$(g.s.$~\rightarrow~$g.s.$)\geq~1.9\times10^{16}$ yr.
 \end{center}

To search for the double electron capture of $^{84}$Sr to the
excited level $2^+$ of $^{84}$Kr, the experimental data were
fitted in the energy interval $(868-896)$ keV
by a Gaussian function (to describe the gamma peak with the energy of
881.6 keV) and a
polynomial function of second degree (to approximate the background, see
Fig. 12, b). The fit gives an area of $S=(-1.2\pm 4.9)$ counts for the double $\beta$ process
searched for, giving no evidence for the
effect ($\lim S=6.9$ counts). Taking into account the detection
efficiency for $\gamma$ quanta with energy 882 keV (5.8\%), we set
the following limit on the process:

\begin{center}
 $T_{1/2}^{2\nu 2\varepsilon}$(g.s.$~\rightarrow~881.6$ keV$)\geq 3.1\times 10^{16}$ yr.
\end{center}

In the neutrinoless $2\varepsilon$ capture to the $2^+$ level, two
$\gamma$ quanta should be emitted. The interaction of the additional $\simeq
0.9$ MeV $\gamma$ quantum with the HPGe detector slightly
decreases the efficiency for the 882 keV peak (5.0\%) leading to
the limit:

\begin{center}
 $T_{1/2}^{0\nu 2\varepsilon}$(g.s.$~\rightarrow~881.6$ keV$)\geq 2.6\times 10^{16}$ yr.
\end{center}

All the half-life limits on $2\beta$ decay processes in $^{84}$Sr,
obtained in the present experiment, are summarized in Table~4.
Previously, only one limit on $0\nu\varepsilon\beta^+$ mode was known;
it was derived in \cite{DBD-tab} on the basis of the data of an old
experiment with photoemulsions \cite{Fremlin:1952}, and is two orders
of magnitude lower than the one obtained in this work.
It should be also noted that an experiment to search for $2\beta$ decays in
$^{84}$Sr with SrCl$_2$ crystal scintillator ($\oslash 2 \times 1.5$ cm)
with $4\pi$ CsI(Tl) active shielding is in progress in the Yang-Yang underground
laboratory \cite{Roo08} which has a potential to improve the limits presented here.

\begin{table*}[ht]
\caption{Half-life limits on 2$\beta$ processes in $^{84}$Sr. The
energies of the $\gamma$ lines ($E_\gamma$), which were used to
set the $T_{1/2}$ limits, are listed in column 4 with the
corresponding detection efficiencies ($\eta$) in column 5. The $T_{1/2}$ limits
are derived in the present work at 90\% C.L., while the limit from
\cite{DBD-tab,Fremlin:1952} is given at 68\% C.L.}
\begin{center}
\resizebox{1.00\textwidth}{!}{
\begin{tabular}{lllllll}

 \hline
 Process                   & Decay       & Level of      & $E_\gamma$  & $\eta$ & \multicolumn{2}{c}{$T_{1/2}$ (yr)}  \\
 \cline{6-7}
 of decay                  & mode        & daughter      & (keV)       &        & Present work        & \cite{DBD-tab,Fremlin:1952}  \\
 ~                         & ~           & nucleus       &             &        & ~                   & ~                            \\
 ~                         & ~           & (keV)         &             &        &                     & ~                            \\
 \hline
 ~                         & ~           & ~             & ~           & ~      & ~                   & ~                            \\
 $\varepsilon\beta^+$      & $0\nu$      & g.s.          & 511         & 7.2\%  & $>6.9\times10^{15}$ & $>7.3\times10^{13}$          \\
 $\varepsilon\beta^+$      & $2\nu$      & g.s.          & 511         & 7.2\%  & $>6.9\times10^{15}$ & --                           \\
 $2K$                      & $0\nu$      & g.s.          & $1754-1762$ & 4.0\%  & $>6.0\times10^{16}$ & --                           \\
 $KL$                      & $0\nu$      & g.s.          & $1767-1775$ & 3.9\%  & $>1.9\times10^{16}$ & --                           \\
 $2L$                      & $0\nu$      & g.s.          & $1779-1788$ & 3.9\%  & $>5.9\times10^{16}$ & --                           \\
 $2\varepsilon$            & $0\nu$      & $2^+$ 881.6   & 881.6       & 5.0\%  & $>2.6\times10^{16}$ & --                           \\
 $2\varepsilon$            & $2\nu$      & $2^+$ 881.6   & 881.6       & 5.8\%  & $>3.1\times10^{16}$ & --                           \\
 ~                         & ~           & ~             & ~           & ~      & ~                   & ~                            \\
 \hline
 \end{tabular}
 }
 \end{center}
 \end{table*}

\section{Conclusions}

The radioactive contamination of the SrI$_2$(Eu) crystal scintillator
obtained using a Stockbarger growth technique was estimated with
the help of two approaches: by low background measurements in
scintillation mode at sea level, and with the help of ultra-low background HPGe
$\gamma$ ray spectrometry deep underground. We have found a contamination of the
scintillator by $^{137}$Cs, $^{226}$Ra and $^{228}$Th on the level of 0.05
Bq/kg, 0.1 Bq/kg and 0.01 Bq/kg, respectively. Only limits were set on the
contamination of the detector by $^{138}$La at level of $\leq 0.02$ Bq/kg,
while the activities of $^{40}$K, $^{90}$Sr, $^{152}$Eu, $^{154}$Eu,
$^{176}$Lu are below the detection limits
of $(0.1-0.3)$ Bq/kg. The intrinsic radiopurity of the
SrI$_2$(Eu) scintillator is still far from NaI(Tl) and CsI(Tl)
scintillators developed for low counting experiments, while it is
three orders of magnitude better than that of the
scintillation materials containing La, and five orders of
magnitude better than that of the scintillators containing Lu.

The response of the SrI$_2$(Eu) crystal scintillator to $\alpha$
particles was estimated by using the trace contamination of the
crystal by $^{226}$Ra. The $\alpha/\beta$ ratio was measured as
0.55 for 7.7 MeV $\alpha$ particles of $^{214}$Po. No difference
in pulse shapes of scintillation for $\gamma$ quanta and
$\alpha$ particles was observed (the decay time was estimated to be: $\approx 1.7~\mu$s).

Applicability of SrI$_2$(Eu) crystal scintillators to the search for the 
double beta decay of $^{84}$Sr was demonstrated for the first
time. New improved half-life limits were set on double electron
capture and electron capture with positron emission in $^{84}$Sr
at level of $T_{1/2}\sim 10^{15}-10^{16}$ yr.

The results of these studies demonstrate the possible perspective of
the SrI$_2$(Eu) highly efficient scintillation material in a variety of
applications, including low counting measurements.

An R\&D of SrI$_2$(Eu) crystal scintillators is in progress. We are going to study 
radioactive contaminations of larger volume SrI$_2$(Eu) crystal scintillators 
both by ultra-low background HPGe $\gamma$ spectrometry and low background scintillation counting at the Gran Sasso National Laboratory.

\section{Acknowledgments}

The work of the INR Kyiv group was supported in part by the
Project ``Kosmomikrofizyka-2'' (Astroparticle Physics) of the
National Academy of Sciences of Ukraine.

\end{document}